\documentclass[runningheads]{llncs}
\usepackage[T1]{fontenc}
\usepackage{graphicx}
\usepackage{tabularx}
\usepackage{makecell}
\usepackage{multirow}
\usepackage{enumitem}
\usepackage{amsmath}
\usepackage{orcidlink}
\begin{document}
\title{Whole Slide Multiple Instance Learning for Predicting Axillary Lymph Node Metastasis} 
\titlerunning{Whole Slide Multiple Instance Learning}
%\author{***}
\author{Glejdis Shkëmbi\inst{1} \and
Johanna P. Müller\inst{1}\orcidlink{0000-0001-8636-7986} \and
Zhe Li\inst{1} \and
Katharina Breininger \inst{1} \orcidlink{0000-0001-7600-5869}\and
Peter Schüffler \inst{2} \orcidlink{0000-0002-1353-8921}\and
Bernhard Kainz \inst{1, 3}\orcidlink{0000-0002-7813-5023}}
\authorrunning{Shkëmbi et al.}
\institute{Friedrich-Alexander-Universität Erlangen-Nürnberg, Germany \email{glejdisshkembi@gmail.com, johanna.paula.mueller@fau.de} \and Technical University of Munich, Germany \and Imperial College London, United Kingdom \\}
%Springer Heidelberg, Tiergartenstr. 17, 69121 Heidelberg, Germany

%
\maketitle              % typeset the header of the contribution
\begin{abstract}
Breast cancer is a major concern for women's health globally, with axillary lymph node (ALN) metastasis identification being critical for prognosis evaluation and treatment guidance. This paper presents a deep learning (DL) classification pipeline for quantifying clinical information from digital core-needle biopsy (CNB) images, with one step less than existing methods. A publicly available dataset of $1058$ patients was used to evaluate the performance of different baseline state-of-the-art (SOTA) DL models in classifying ALN metastatic status based on CNB images. An extensive ablation study of various data augmentation techniques was also conducted. Finally, the manual tumor segmentation and annotation step performed by the pathologists was assessed. Our proposed training scheme outperformed SOTA by $3.73\%$. Source code is available \href{https://github.com/glejdis/Whole-Slide-MIL-for-Predicting-Axillary-Lymph-Node-Metastasis}{here}.

\keywords{Breast cancer  \and Auxillary lymph node\and Multiple instance learning \and Tumour segmentation.}
\end{abstract}

\section{Introduction}
Breast cancer (BC) is currently the world’s most diagnosed cancer~\cite{Ferlay_2021}, accounting for $685,000$ deaths among women in $2020$. Axillary lymph nodes (ALNs) are typically the first location of breast cancer metastasis, which makes their status the single most important predictive indicator for diagnosis \cite{Jatoi_1999}. Several studies have demonstrated that deep learning (DL) can support pathologists by increasing the sensitivity of ALN micro-metastasis detection, as shown in the review paper \cite{steiner_2018}.
Furthermore, earlier research has demonstrated lymph node metastasis detection may be assisted by deep features from whole slide images (WSIs)~\cite{Zhao_2020}. 
Deep learning has greatly reduced the need for domain experts to manually extract features from data. However, human experts are still essential for data labeling. Unfortunately, the increasing complexity of many problems requires large amounts of annotated data, which can be costly and raise privacy concerns since the process of labeling often involves analyzing and categorizing personal information or sensitive data, especially, in a medical context.
Pixel-level annotation is a time-consuming and expensive process, as it requires a human annotator to go through every single pixel in an image to label it correctly. This process can take a lot of time and effort, especially for large datasets or complex images, and may also require a high level of expertise or specialized training. On the other hand, image-level annotation is a much simpler and faster process, as it only requires a human annotator to assign a single label to an entire image.

Multiple Instance Learning (MIL) has enabled pathologists to label bags of sub-images or patches, rather than labeling each individual patch. This approach is particularly useful in binary classification tasks, where distinguishing between healthy and diseased patients is the primary goal. By using MIL, an image can be labeled as malignant if it contains at least one malignant patch, while an image is considered cancer-free if all patches are classified as healthy.
We present a pipeline that enables the prediction of axillary lymph node metastasis status from whole slide images of core-needle biopsy samples from patients with early breast cancer. The prediction of metastasis status based solely on histopathological images is a complex and challenging task and one that previously exceeded the competencies of medical experts. Our approach surpasses the limitations of traditional manual assessments and pixel-wise annotation of whole slide images, but yet provides a reliable and objective method for predicting metastasis status.

In this paper, we reproduce the results of the attention-based MIL classification model proposed by \cite{Xu_2021}, which aims at identifying the (micro-)metastatic status of ALN preoperatively using the Early Breast Cancer Core-Needle Biopsy WSI (BCNB) \cite{Xu_2021} dataset of early breast cancer (EBC) patients. To evaluate the impact of feature extraction components on the classification pipeline's performance, we utilized $43$ different convolutional neural networks (CNNs) as a backbone for feature extraction. 
Given the constraints of limited data, we hypothesize that data augmentation can enhance and diversify the original BCNB dataset \cite{Xu_2021}, leading to improved model generalization and overall performance. An extensive ablation study on different data augmentation techniques, including basic and advanced methods, examined this effect. 
Finally, the pipeline proposed by~\cite{Xu_2021} relied on hand-labeled, pixel-wise annotation of the tumor. However, manual tumor segmentation and annotation is a time-consuming process and prone to errors. As such, we tested the necessity of this information input for the deep-learning core-needle biopsy model.

\noindent\textbf{Related Work.} Recently, DL has demonstrated its ability to extract features from medical images at a high throughput rate and analyze the correlation between primary tumor features and ALN metastasis information. In the study by \cite{Qiang_2020}, researchers utilized Inception-v3 \cite{Szegedy_2015}, Inception-ResNet-v2 and ResNet \cite{He_2015} to predict clinically negative ALN metastasis using two-dimensional grey-scale ultrasound images of patients with primary breast cancer. Moreover, \cite{Zheng_2020} employed two-dimensional shear wave elastography (SWE), a new ultrasound method for measuring tissue stiffness, to discriminate between malignant and benign breast tumors. They combined clinical parameters with DL radiomics, where a pre-trained ResNet model \cite{He_2015} encoded the input images into features. This combination achieved an area under the receiver operating characteristic (AUROC) value of $0.902$ in the test cohort. Comparatively, their classification model using only clinicopathologic data achieved an AUROC value of $0.727$.

\section{Method}
Our goal in this paper is to predict the status of axillary lymph nodes (ALN) using core needle biopsies. We present the deep learning (DL)-based core-needle biopsy with whole slide processing (DLCNBC-WS) network, a revised method that builds on two attention-based deep MIL frameworks, proposed by~\cite{Xu_2021}. The Deep-learning core-needle biopsy (DLCNB) model is the base model for both frameworks and is built on an attention-based deep MIL framework for predicting the ALN status using DL (histopathological) features~\cite{Ilse_2018}, from core-needle biopsies. These DL features were extracted only from the cancerous areas of the WSIs of breast CNB samples. We differentiate between non-existing or already advanced metastasis as ALN status. In contrast to the DLCNB model, the DLCNBC model uses additionally selected features from the clinical data set. The clinical information of the slide is added to all the constructed bags in order to provide insightful information for predicting and achieving better performance. The whole algorithm pipeline is composed of four steps and follows~\cite{Xu_2021}.

One of the key features of our revised method is the eliminated segmentation of the cancerous features in the CNB slides which allows higher flexibility and adaptability. Our proposed DLCNBC with whole slide processing (DLCNBC-WS) omits the time-consuming step of segmentation and reduces the requirements of its applications. We accomplish to cut the steps of the algorithm pipeline into only three remaining steps.

\begin{figure}[h!]
    \centering
    \includegraphics[width=1\columnwidth]{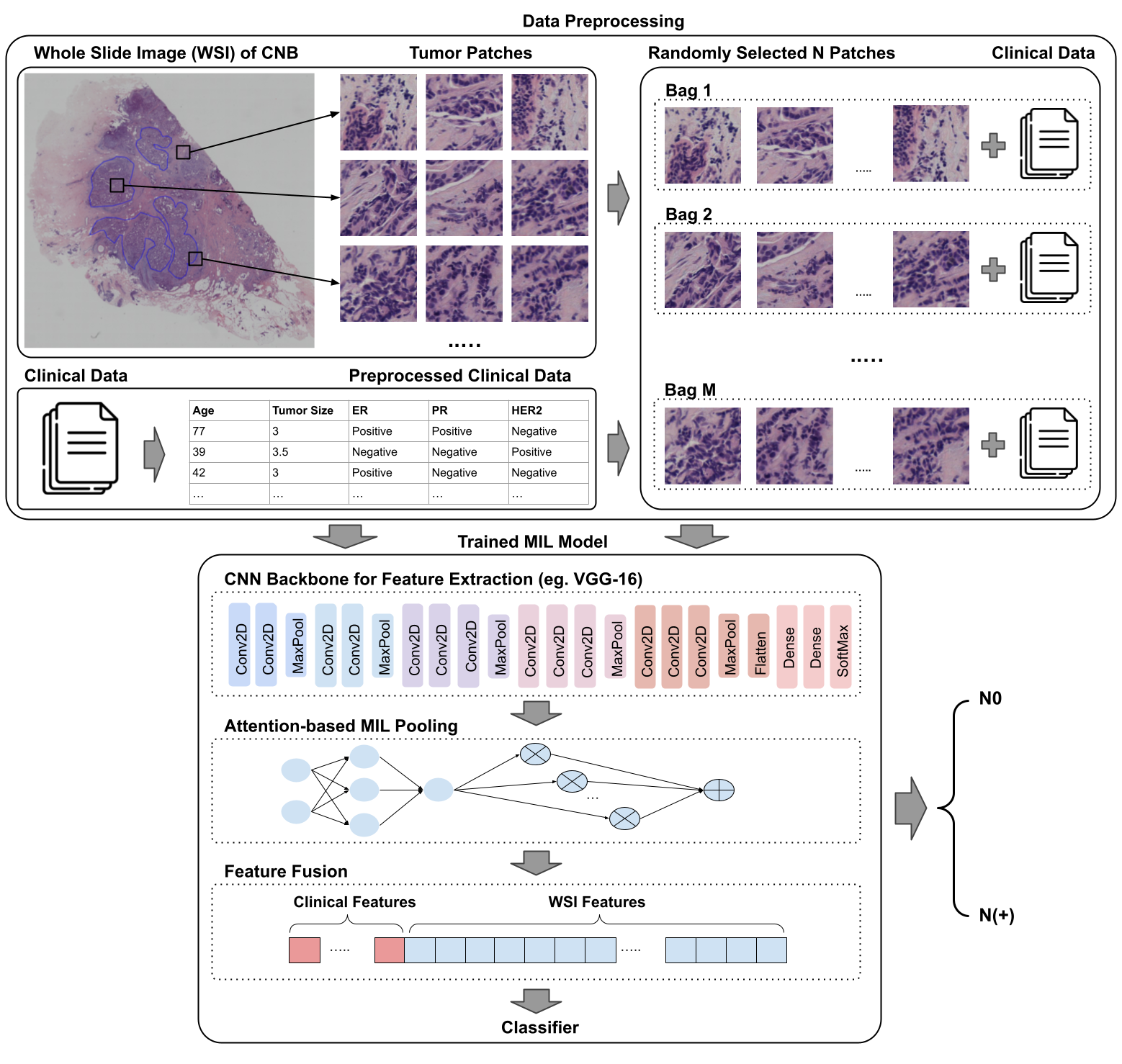}
    \caption[The overall pipeline of DLCNBC-WS.]{The overall pipeline of DLCNBC-WS model to predict ALN status between N0 and N(+). Top image: Multiple training bags were built. Bottom image: DLCNBC-WS model training process included feature extraction and MIL, and by aggregating the model outputs of all bags from the same slide, the ALN state is predicted.
    }
    \label{fig:DLCNBC}
\end{figure}

In the following, we explain the DLCNBC-WS in detail, see Fig.~\ref{fig:DLCNBC}. First, $N$ feature vectors for the $N$ WSI patches of size $256 \times 256$ pixels in each bag were extracted using multiple different CNN models as the backbone for feature extraction. Multiple bags were built for each WSI (top image of Fig.~\ref{fig:DLCNBC}). As the bottom image of Fig.~\ref{fig:DLCNBC} shows, the clinical data were also preprocessed for feature extraction in addition to the WSIs. For tackling the restricted availability of the training data and therefore the issue of overfitting, a wide range of basic and advanced data augmentation techniques were applied to the WSI patches. 

Next, the $N$ feature vectors of patches in a bag were processed by max-pooling and reshaping. Then, they were passed through two fully connected layers to output $N$ weight factors for patches in the bag, which were then further multiplied by the original feature vectors in order to dynamically adjust the impact of instance features. Finally, the weighted image feature vectors, created by the aggregation of features of multiple patches from the attention module, were fused with clinical features by concatenation. A classifier learns to estimate ALN status based on aggregated input.

\section{Evaluation}

\noindent\textbf{Data Source.} 
\label{data_sources_prerpocess}
The \href{https://bcnb.grand-challenge.org/Dataset/}{BCNB} dataset used in this paper includes clinical data and the core-needle biopsy (CNB) hematoxylin and eosin (H\&E) stained WSIs of patients with pathologically confirmed ALN status. It includes the following clinical features: age, tumor size, tumor type, estrogen receptor (ER)-status, progesterone receptor (PR)-status, human epidermal growth factor receptor $2$ (HER2)-status, HER2 expression, histological grading, surgical procedure (ALND or SLNB), Ki67, molecular subtype, the number of lymph node metastases (LNM) and the target variable which is the metastatic status of ALN (N0, N+(1-2) or N+($>$2)). The dataset contains $655$ patients with N0 status (no metastasis), $210$ patients with N+(1-2) status (cancer has metastasized to $1$-$2$ ALNs), and $193$ patients with N+($>$2) ALN status (cancer has metastasized to $3$ or more ALNs). 

\noindent\textbf{Preprocessing.} 
The selected numerical features (age and tumor size) of the clinical dataset were standardized and the selected categorical variables (ER/ PR/ HER-2 status and the ALN metastases status as label) were one-hot encoded as no ordinal relationship was observed. The WSI preparation was performed according to \cite{Xu_2021}. The multiple training bags were constructed by fusing together the features extracted from the cropped patches of the selected tumor regions of each CNB WSI with the selected features from the clinical dataset.
The WSIs and the clinical data were divided on a per-patient level into training and independent testing cohorts; 80:20 respectively, according to \cite{Xu_2021}, where one WSI belongs to one patient and clinical information of each patient is represented by one entry in the clinical dataset. The validation cohort was created from the training set by randomly selecting $25 \%$ of that data. In this way, the patches from one patient are found exclusively in either the train/validation set or the test set, but not in both. 
In \cite{Xu_2021}, the model relied on manually annotated cancerous regions in each WSI. However, to reduce the need for expert pathologists, decrease human errors in annotation, and increase reproducibility, patches were sampled from all (tissue) areas of the WSI, in contrast to the pipeline by \cite{Xu_2021}. They extracted the patches from only tumor areas which were manually selected in advance by pathologists. Our pipeline framework assumes that the model has no prior knowledge of whether the sampled patch belongs to a tumor. For adapting the dataset to our proposed pipeline, the sampled patches with artefacts or without cell information were filtered out using Shannon entropy computed on the grey-scale versions of the patches. 

\noindent\textbf{Training.} 
The model is trained to predict the ALN metastasis based on WSIs of primary BC samples by predicting the bag label while thoroughly considering all included instances in each bag. A stochastic gradient descent (SGD) optimizer with a learning rate of $1e^{-4}$ to update the model parameters and a cosine annealing warm restarts strategy to adjust the learning rate were utilized. In the training phase, L2 regularization was implemented by passing weight decay of $1e^{-3}$ to the optimizer. In addition, to ease overfitting L1 regularization was employed with weight $1e^{-6}$.  During the testing phase, the ALN status was predicted by aggregating the model outputs of all bags from the same WSI (Fig.~\ref{fig:DLCNBC}c). We trained our models on NVIDIA RTX A6000 for $2000$ epochs.

\noindent\textbf{Statistical Analysis.}
To examine the clinical characteristics of different cohorts, we conducted a correlation analysis. To establish a basis for comparison, we first utilized all features of the clinical dataset to develop a logistic regression (LR) model for ALN prediction. We then trained the LR model using a subset of features from the clinical dataset, specifically age, tumor size, and the expression levels of ER, PR, and HER2 biomarkers. These specific features were chosen based on the results of a correlation analysis conducted on the clinical dataset, which involved selecting features with a higher correlation to the outcome variable and a lower correlation among other predicting variables. The reduced clinical dataset was then incorporated into the input of the DLCNBC model.

\noindent\textbf{Ablation Study.}
Comparing multiple backbone models is essential to building a robust and effective deep-learning classification pipeline. We compared multiple CNN models for finding the best-suited pipeline for this specific task. We updated the list of backbone models with the latest advances: AlexNet, VGG with and without BN, GoogleNet, Inception-v3, ResNet, DenseNet, SqueezeNet, ResNext, WideResNet, MobileNet, ShuffleNet-v2, EfficientNet, and EfficientNet-v2. 

To identify which augmentation techniques contribute the most to improving the model's accuracy and generalization ability, we conducted an extensive ablation study. All augmentations techniques were applied as a preprocessing step on the training cohort of the WSI patches. For model training, each image was resized to $224 \times 224$ pixels and normalized. Next, different data augmentation methods, including different types of geometric augmentation (including flipping, rotation, translation, shearing and scaling), colors space augmentation (including changes in brightness, contrast, saturation and hue, converting an image to grayscale, solarization, and posterization), image erasing (including random erasing and random cropping), AugMix \cite{Hendrycks_2019}, as well as more advanced approaches such as AutoAugment \cite{Cubuk_2019}, RandAugment \cite{Cubuk_2020} and TrivialAugmentWide \cite{Müller_2021}, were applied. Basic augmentation techniques such as geometric augmentation, color space augmentation, image erasing and image mixing were manually designed, and a manually predefined set of parameter values was applied. The augmentation techniques that yielded the best performance on the test dataset, as determined by AUROC, were selected and combined into an augmentation mixture, which was subsequently evaluated to assess its impact on model performance. The best-performing model based on~\cite{Xu_2021} (DLCNBC model with VGG-16 BN) was trained on each data augmentation technique. 
Finally, the DLCNBC and DLCNBC-WS models were trained, as described above.

\section{Results and Discussion}

In this section, we present and analyze the results of our experiments. First, the performance results in terms of AUROC, accuracy, sensitivity, specificity, PPV, NPV, and F1-score in the test cohort of the best and worst-performing DLCNBC models with different backbones for feature extraction in the binary classification of ALN status are illustrated in Table \ref{best_worst_DLCNBC}. The highest AUROC score of $0.837$ was achieved using VGG-13 with batch normalization (BN) as the feature extractor, while the lowest AUROC score of $0.543$ was achieved by EfficientNet-b0, EfficientNet-b2 and EfficientNet-b3. In general, EfficientNet and Inception-v3 have shown not to outperform VGG in terms of accuracy and AUROC while using fewer computational resources. However, VGG outperformed all other backbone models, for which its reasons needs to be further investigated in future.
Second, the ablation study showed that the classification was improved by further applying one of the following augmentation strategies to the data for DLCNBC models: random rotation, scaling, shearing, or random vertical flipping, see Tab.~\ref{augmentation}. The best results were obtained by performing a random rotation $\leq10^{\circ}$. 
It increased the AUROC score from $0.837$ to $0.852$ in the binary classification task. 
\begin{table}[h!]
\caption{The performance results on the test cohort of the best and worst performing DLCNBC models with different backbones for feature extraction in the binary classification of ALN status (N0 vs. N(+)).}
\label{best_worst_DLCNBC}
\centering
        \begin{tabular}{lcccccccccc}
        \toprule
        {Backbone } & AUROC & Accuracy & Sensitivity&Specificity&PPV &NPV&F1-Score\\
         \cmidrule(lr){1-1}\cmidrule(lr){2-2}\cmidrule(lr){3-3}\cmidrule(lr){4-4}\cmidrule(lr){5-5}\cmidrule(lr){6-6}\cmidrule(lr){7-7}\cmidrule(lr){8-8}
        VGG-11 BN  & $0.83$ & $\mathbf{0.739}$ & $0.607$ & $0.821$ & $0.68$ & $0.769$ & $0.642$  \\
        VGG-13 BN & $\mathbf{0.837}$ & $\mathbf{0.739}$ & $0.548$ & $0.858$ & $0.708$ & $0.752$ & $0.617$ \\
        VGG-16 BN & $0.822$ & $\mathbf{0.739}$ & $0.679$ & $0.776$ & $0.655$ & $0.794$ & $0.667$ \\
        VGG-19 BN & $0.828$ & $\mathbf{0.739}$ & $\mathbf{0.881}$ & $0.649$ & $0.612$ & $\mathbf{0.897}$ & $\mathbf{0.722}$ \\
        Inception-v3 & $0.545$ & $0.633$ & $0.095$ & $0.970$ & $0.667$ & $0.631$ & $0.167$ \\
        EfficientNet-b0 & $0.543$ & $0.624$ & $0.024$ & $\mathbf{1.000}$ & $\mathbf{1.000}$ & $0.620$ & $0.047$ \\
        EfficientNet-b2 & $0.545$ & $0.615$ & $0.048$ & $0.970$ & $0.500$ & $0.619$ & $0.087$ \\
        EfficientNet-b3 & $0.543$ & $0.615$ & $0.071$ & $0.955$ & $0.500$ & $0.621$ & $0.125$ \\
         \bottomrule
        \end{tabular}
\end{table}
\begin{table}[h!]
\caption{The performance results on the test cohort of the four best and worst performing data augmentation techniques on the DLCNBC models with VGG-16 BN as the backbone for feature extraction in the binary classification of ALN status (N0 vs. N(+)).}
\label{augmentation}
\centering
        \begin{tabular}{lcccccccccc}
        \toprule
        {Augmentation} & AUROC & Accuracy & Sensitivity&Specificity&PPV &NPV&F1-Score\\
         \cmidrule(lr){2-2}\cmidrule(lr){3-3}\cmidrule(lr){4-4}\cmidrule(lr){5-5}\cmidrule(lr){6-6}\cmidrule(lr){7-7}\cmidrule(lr){8-8}
        Random Rotation  & $\mathbf{0.868}$ & $0.761$ & $0.690$ & $0.806$ & $0.690$ & $0.806$ & $0.690$  \\
        Shear  & $0.865$ & $0.775$ & $0.679$ & $\mathbf{0.836}$ & $\mathbf{0.722}$ & $0.806$ & $0.699$  \\
        Random Erasing  & $0.863$ & $0.757$ & $\mathbf{0.821}$ & $0.716$ & $0.645$ & $\mathbf{0.865}$ & $0.723$  \\
        Vertical Flip  & $0.857$ & $\mathbf{0.780}$ & $0.762$ & $0.791$ & $0.696$ & $0.841$ & $\mathbf{0.727}$  \\
         \bottomrule
        \end{tabular}
\end{table}
\begin{table}[h!]
\caption{Comparison of the performance of DLCNBC and DLCNBC-WS models with VGG-13 BN backbone for binary classification of the ALN status. T: training cohort, V: validation cohort, I-T: independent test cohort.}
\label{vgg13bn_comaparison}
\centering
        \begin{tabular}{lcccccccccc}
        \toprule
        {Method } & Set & AUROC & Accuracy & Sensitivity&Specificity&PPV &NPV&F1-Score\\
         \cmidrule(lr){2-2}\cmidrule(lr){3-3}\cmidrule(lr){4-4}\cmidrule(lr){5-5}\cmidrule(lr){6-6}\cmidrule(lr){7-7}\cmidrule(lr){8-8}\cmidrule(lr){9-9}
        DLCNBC  & T &  $0.903$ & $0.814$ & $0.812$ & $0.815$ & $0.73$ & $0.876$ & $0.769$ \\
        & V &  $0.847$ &  $0.767$& $0.734$ & 0.786& 0.674 & 0.831 & 0.703 \\
        & I-T &   $0.837$ & $0.739$ & $0.548$ & $\mathbf{0.858}$ & $0.708$ & $0.752$ & $0.617$  \\
        \midrule
        DLCNBC & T &  $0.945$ & $0.881$ & $0.896$ & $0.872$ & $0.811$ & $0.932$ & $0.851$ \\
        Rotation  & V & $0.866$ & $0.790$ & $0.886$ & $0.733$ & $0.667$ & $0.914$ & $0.761$  \\
         & I-T &  $0.852$ & $0.766$ & $0.774$ & $0.761$ & $0.670$ & $0.843$ & $0.718 $  \\        \midrule
        DLCNBC-WS & T &  $0.983$ & $0.938$ & $0.941$ & $0.935$ & $0.899$ & $0.963$ & $0.920$     \\
        & V &  $0.863$ & $0.781$ & $0.823$ & $0.756$ & $0.670$ & $0.876$ & $0.739$  \\
        & I-T &   $\mathbf{0.862}$ & $\mathbf{0.803}$ & $\mathbf{0.833}$ & $0.784$ & $0.707$ & $\mathbf{0.882}$ & $\mathbf{0.765}$ \\   \midrule
        DLCNBC-WS & T &   $0.976$ & $0.910$ & $0.907$ & $0.912$ & $0.863$ & $0.941$ & $0.885$     \\
        Rotation & V &  $0.884$ & $0.790$ & $0.709$ & $0.840$ & $0.727$ & $0.827$ & $0.718$ \\
        & I-T &  $0.843$ & $0.766$ & $0.619$ & $\mathbf{0.858}$ & $\mathbf{0.732}$ & $0.782$ & $0.671$   \\
         \bottomrule
        \end{tabular}
\end{table}

\begin{table}[h!]
\caption{The performance results on the test cohort of the DLCNBC-WS models with different backbones for feature extraction in the binary classification of ALN status (N0 vs. N(+)).}
\label{best_worst_DLCNBCWC_binary}
\centering
        \begin{tabular}{lcccccccccc}
        \toprule
        {Backbone } & AUROC & Accuracy & Sensitivity&Specificity&PPV &NPV&F1-Score\\
         \cmidrule(lr){2-2}\cmidrule(lr){3-3}\cmidrule(lr){4-4}\cmidrule(lr){5-5}\cmidrule(lr){6-6}\cmidrule(lr){7-7}\cmidrule(lr){8-8}
        VGG-11 & $0.82$ & $0.711$ & $0.655$ & $0.746$ & $0.618$ & $0.775$ & $0.636$  \\
        VGG-11 BN & $\mathbf{0.86}$ & $0.739$ & $0.940$ & $0.612$ & $0.603$ & $\mathbf{0.943}$ & $0.735$  \\
        VGG-13 & $0.84$ & $0.78$ & $0.690$ & $0.836$ & $0.725$ & $0.812$ & $0.707$ \\
        VGG-13 BN & $\mathbf{0.86}$ & $\mathbf{0.803}$ & $\mathbf{0.833}$ & $0.784$ & $0.707$ & $0.882$ & $\mathbf{0.765}$ \\
        VGG-16 & $0.85$ & $0.766$ & $0.571$ & $0.888$ & $\mathbf{0.762}$ & $0.768$ & $0.653$ \\
        VGG-16 BN & $\mathbf{0.86}$ & $0.706$ & $0.357$ & $\mathbf{0.925}$ & $0.750$ & $0.697$ & $0.484$ \\
        VGG-19 & $0.83$ & $0.784$ & $0.738$ & $0.813$ & $0.713$ & $0.832$ & $0.725$ \\
        VGG-19 BN & $0.85$ & $0.766$ & $0.571$ & $0.888$ & $\mathbf{0.762}$ & $0.768$ & $0.653$ \\
         \bottomrule
        \end{tabular}
\end{table}

\begin{table}[h!]
\caption{The performance results for each class on the test cohort of the best and worst performing DLCNBC models with different backbones for feature extraction in the multi-class classification of ALN status (N0 vs. N+(1-2) vs. N+($>$2)).}

\label{best_worst_DLCNBC_multiclass}
\centering
        \begin{tabular}{lcccccccccc}
        \toprule
        {Backbone } & Class & AUROC & Accuracy & Sensitivity&Specificity&PPV &NPV\\
         \cmidrule(lr){1-1}\cmidrule(lr){2-2}\cmidrule(lr){3-3}\cmidrule(lr){4-4}\cmidrule(lr){5-5}\cmidrule(lr){6-6}\cmidrule(lr){7-7}\cmidrule(lr){8-8}
         VGG-11 BN & N0 & 0.84 & $0.761$ & $0.855$ & $0.621$ & $\mathbf{0.772}$ & $0.740$ \\
         & N+(1-2) & $0.80$ & $0.784$ & $0.440$ & $0.887$ & $0.537$ & $0.842$ \\
         & N+($>$2) & $0.68$ & $0.766$ & $0.243$ & $0.873$ & $0.281$ & $0.849$ \\
         \midrule
         VGG-13 BN & N0 & $0.84$ & $0.757$ & $0.847$ & $0.621$ & $0.771$ & $0.730$  \\
         & N+(1-2) & $0.79$ & $0.784$ & $0.420$ & $0.893$ & $0.538$ & $0.838$  \\
         & N+($>$2) & $0.71$ & $0.761$ & $0.270$ & $0.862$ & $0.286$ & $\mathbf{0.852}$  \\
         \midrule
         VGG-16 BN & N0 & $\mathbf{0.85}$ & $0.748$ & $0.931$ & $0.471$ & $0.726$ & $0.820$  \\
         & N+(1-2) & $0.79$ & $0.784$ & $0.360$ & $0.911$ & $0.545$ & $0.827$  \\
         & N+($>$2) & $0.71$ & $0.807$ & $0.162$ & $0.939$ & $0.353$ & $0.846$  \\
         \midrule
         WideResNet-101-2 & N0 & $0.58$ & $0.587$ & $0.939$ & $0.057$ & $0.600$ & $0.385$  \\
         & N+(1-2) & $0.60$ & $0.775$ & $0.060$ & $\mathbf{0.988}$ & $0.600$ & $0.779$  \\
         & N+($>$2) & $0.48$ & $0.794$ & $0.000$ & $0.956$ & $0.000$ & $0.824$ \\
         \midrule
         MobileNet v3 large & N0 & $0.64$ & $0.615$ &$ \mathbf{0.947}$ & $0.115$ & $0.617$ & $0.588$  \\
         & N+(1-2) &$ 0.60$ & $0.775$ & $0.120$ & $0.970$ & $0.545$ & $0.787$ \\
         & N+($>$2) &$ 0.50 $& $\mathbf{0.812}$ & $0.027$ & $0.972$ & $0.167$ & $0.830$ \\
         \midrule
         EfficientNet b3 & N0 & $0.63$ &$ 0.606$ & $\mathbf{0.947}$ & $0.092$ & $0.611$ & $0.533 $ \\
         & N+(1-2) & $0.59$ & $0.780$ & $0.120$ & $0.976$ & $0.600$ & $0.788$  \\
         & N+($>$2) & $0.51$ & $0.807$ & $0.000$ &$ 0.972$ & $0.000$ & $0.826$  \\
         \midrule
         EfficientNet b5 & N0 & $0.64$ & $0.601$ & $0.916$ & $0.126$ & $0.612$ & $0.500 $ \\
         & N+(1-2) &$ 0.62 $& $0.752$ & $0.100$ & $0.946$ & $0.357$ & $0.779$  \\
         & N+($>$2) & $0.48 $& $\mathbf{0.812}$ & $0.054$ & $0.967$ & $0.250$ & $0.833$  \\
         \bottomrule
        \end{tabular}
\end{table}

Third, our proposed DLCNBC-WS model performed best when VGG-13 BN was used as the backbone for feature extraction in the binary classification task N0 vs. N(+), see Tab.~\ref{best_worst_DLCNBCWC_binary}, the results for multi-class classification are given in Tab.~\ref{best_worst_DLCNBC_multiclass}. 

Tab.~\ref{vgg13bn_comaparison} compares the results of the baseline method DLCNBC and our proposed DLCNBC-WS with the best-performing backbone model for binary classification, VGG-13 with BN. We also show the impact of random rotation as an augmentation function for both models. The table shows that applying random rotation increased the model performance in terms of AUROC score only for the baseline method by $1.79$ $\%$. Moreover, training the model without the expert tumor segmentation step surpassed the results of the preceding paper on the BCNB dataset \cite{Xu_2021} for the independent test cohort by $3.37$ $\%$ in the AUROC score. 

\section{Conclusion}
In this paper, we succeeded in reproducing the results of \cite{Xu_2021} and further upgraded their attention-based MIL classification pipeline for predicting ALN metastasis status preoperatively in EBC patients. We found that both the performance of DLCNB and DLCNBC were significantly influenced by CNN backbone selection. The results of the ablation study showed that learning was heavily influenced by the preparation of training data and changes in data distribution. In particular, the use of random rotation on top of the baseline model outperformed SOTA in the binary classification of ALN status by $2.53$ $\%$. Lastly, the requirement of pathologists to perform tumor segmentation and annotation was removed and SOTA was outperformed by $3.73$ $\%$. 

Since the waiting time between biopsy and pathological classification affects whether a diagnosis of lymph node metastasis still reflects the current status and, hence, is accurate, the ALN metastasis is intrinsically unstable~\cite{Qiang_2020}. For instance, if monitored for a sufficient amount of time, some patients with negative lymph nodes may eventually develop positive lymph nodes. In addition, an interesting attempt would be to evaluate the clinical utility of immunochemically stained images, instead of focusing solely on H\&E staining. One important limitation of CNNs in histopathology image analysis does not clearly capture inter-nuclear interactions and histopathological information. This is important for detecting and characterizing cancers \cite{Jaume_2020}, which could be solved with the help of graph neural networks. 

\noindent\textbf{Acknowledgements}: The authors gratefully acknowledge the scientific support and HPC resources provided by the Erlangen National High Performance Computing Center (NHR@FAU) of the Friedrich-Alexander-Universität Erlangen-Nürnberg (FAU) under the NHR projects b143dc and b180dc. NHR funding is provided by federal and Bavarian state authorities. NHR@FAU hardware is partially funded by the German Research Foundation (DFG) – 440719683. Additional support was also received by the ERC - project MIA-NORMAL 101083647,  DFG KA 5801/2-1, INST 90/1351-1 and by the state of Bavaria.
%\newpage
\bibliographystyle{splncs04}
\bibliography{paper}

\begin{thebibliography}{10}
\providecommand{\url}[1]{\texttt{#1}}
\providecommand{\urlprefix}{URL }
\providecommand{\doi}[1]{https://doi.org/#1}

\bibitem{Cubuk_2019}
Cubuk, E.D., Zoph, B., Mane, D., Vasudevan, V., Le, Q.V.: Autoaugment: Learning
  augmentation policies from data (2018). \doi{10.48550/ARXIV.1805.09501},
  \url{https://arxiv.org/abs/1805.09501}

\bibitem{Cubuk_2020}
Cubuk, E.D., Zoph, B., Shlens, J., Le, Q.V.: Randaugment: Practical automated
  data augmentation with a reduced search space (2019).
  \doi{10.48550/ARXIV.1909.13719}, \url{https://arxiv.org/abs/1909.13719}

\bibitem{Ferlay_2021}
Ferlay, J., Colombet, M., Soerjomataram, I., Parkin, D.M., Piñeros, M., Znaor,
  A., Bray, F.: Cancer statistics for the year 2020: An overview. International
  Journal of Cancer  \textbf{149}(4),  778--789 (2021).
  \doi{https://doi.org/10.1002/ijc.33588},
  \url{https://onlinelibrary.wiley.com/doi/abs/10.1002/ijc.33588}

\bibitem{He_2015}
He, K., Zhang, X., Ren, S., Sun, J.: Deep residual learning for image
  recognition (2015). \doi{10.48550/ARXIV.1512.03385},
  \url{https://arxiv.org/abs/1512.03385}

\bibitem{Hendrycks_2019}
Hendrycks, D., Mu, N., Cubuk, E.D., Zoph, B., Gilmer, J., Lakshminarayanan, B.:
  Augmix: A simple data processing method to improve robustness and uncertainty
  (2019). \doi{10.48550/ARXIV.1912.02781},
  \url{https://arxiv.org/abs/1912.02781}

\bibitem{Ilse_2018}
Ilse, M., Tomczak, J.M., Welling, M.: Attention-based deep multiple instance
  learning. Proceedings of the 35th International Conference on Machine
  Learning (ICML)  \textbf{abs/1802.04712} (2018),
  \url{http://arxiv.org/abs/1802.04712}

\bibitem{Jatoi_1999}
Jatoi, I., Hilsenbeck, S.G., Clark, G.M., Osborne, C.K.: Significance of
  axillary lymph node metastasis in primary breast cancer  (1999).
  \doi{10.1200/JCO.1999.17.8.2334},
  \url{https://pubmed.ncbi.nlm.nih.gov/10561295/}

\bibitem{Jaume_2020}
Jaume, G., Pati, P., Bozorgtabar, B., Foncubierta{-}Rodr{\'{\i}}guez, A.,
  Feroce, F., Anniciello, A.M., Rau, T., Thiran, J., Gabrani, M., Goksel, O.:
  Quantifying explainers of graph neural networks in computational pathology.
  IEEE/CVF Conference on Computer Vision and Pattern Recognition (CVPR)
  \textbf{abs/2011.12646} (2020), \url{https://arxiv.org/abs/2011.12646}

\bibitem{Müller_2021}
Müller, S.G., Hutter, F.: Trivialaugment: Tuning-free yet state-of-the-art
  data augmentation (2021). \doi{10.48550/ARXIV.2103.10158},
  \url{https://arxiv.org/abs/2103.10158}

\bibitem{steiner_2018}
Steiner, D., MacDonald, R., Liu, Y., Truszkowski, P., Hipp, J., Gammage, C.,
  Thng, F., Peng, L., Stumpe, M.: Impact of deep learning assistance on the
  histopathologic review of lymph nodes for metastatic breast cancer. The
  American Journal of Surgical Pathology  \textbf{42}, ~1 (10 2018).
  \doi{10.1097/PAS.0000000000001151}

\bibitem{Szegedy_2015}
Szegedy, C., Vanhoucke, V., Ioffe, S., Shlens, J., Wojna, Z.: Rethinking the
  inception architecture for computer vision (2015).
  \doi{10.48550/ARXIV.1512.00567}, \url{https://arxiv.org/abs/1512.00567}

\bibitem{Xu_2021}
Xu, F., Zhu, C., Tang, W., Wang, Y., Zhang, Y., Li, J., Jiang, H., Shi, Z.,
  Liu, J., Jin, M.: Predicting axillary lymph node metastasis in early breast
  cancer using deep learning on primary tumor biopsy slides. Frontiers in
  Oncology  \textbf{11}, ~4133 (October 2021). \doi{10.3389/fonc.2021.759007},
  \url{https://doi.org/10.3389\%2Ffonc.2021.759007}

\bibitem{Zhao_2020}
Zhao, Y., Yang, F., Fang, Y., Liu, H., Zhou, N., Zhang, J., Sun, J., Yang, S.,
  Menze, B., Fan, X., Yao, J.: Predicting lymph node metastasis using
  histopathological images based on multiple instance learning with deep graph
  convolution. In: 2020 IEEE/CVF Conference on Computer Vision and Pattern
  Recognition (CVPR). pp. 4836--4845 (2020). \doi{10.1109/CVPR42600.2020.00489}

\bibitem{Zheng_2020}
Zheng, X., Yao, Z., Huang, Y., Yu, Y., Wang, Y., Liu, Y., Mao, R., Li, F.,
  Xiao, Y., Wang, Y., Hu, Y., Yu, J., Zhou, J.: Deep learning radiomics can
  predict axillary lymph node status in early-stage breast cancer. Nature
  Communications  \textbf{11} (03 2020). \doi{10.1038/s41467-020-15027-z}

\bibitem{Qiang_2020}
Zhou, L.Q., Wu, X.L., Huang, S.Y., Wu, G.G., Ye, H.R., Wei, Q., Bao, L.Y.,
  Deng, Y.B., Li, X.R., Cui, X.W., Dietrich, C.F.: Lymph node metastasis
  prediction from primary breast cancer us images using deep learning.
  Radiology  \textbf{294}(1),  19--28 (2020). \doi{10.1148/radiol.2019190372},
  pMID: 31746687

\end{thebibliography}

\end{document}